\documentclass[12pt]{article}

\usepackage{epsf}
\usepackage{times}

\begin{document}

\def\btdplot#1#2{\leavevmode\epsfxsize=#2\textwidth \epsfbox{#1}}

\title{Monte Carlo study of the scattering error of a quartz
reflective absorption tube}

\author{Jacek Piskozub \\Institute of Oceanology PAS \\ Powsta\'{n}c\'{o}w
Warszawy 55 81-712 Sopot, Poland \\ piskozub@iopan.gda.pl
\\ \\
Piotr J. Flatau \\ Scripps Institution of Oceanography, University
of California,  San Diego \\La Jolla, CA 92130-0221
\\ pflatau@ucsd.edu
\\ \\
J. V. Ronald Zaneveld \\ College of Oceanic and Atmospheric
Sciences, Oregon State University \\ 104 Ocean Admin Bldg
Corvallis, OR 97331-5503}

\maketitle

In Press: Journal of Atmospheric and Oceanic Technology, 2000

\abstract{A Monte Carlo model was used to study the scattering
error of an absorption meter with a divergent light beam and a
limited acceptance angle of the receiver. Reflections at both ends
of the tube were taken into account. Calculations of the effect of
varying optical properties of water, as well as the receiver
geometry, were performed. A weighting function showing the
scattering error quantitatively as a function of angle was
introduced.  Some cases of the practical interests are discussed}

\section{INTRODUCTION}

Light absorption is the essential element required for marine phytoplankton growth. The
region of the upper ocean illuminated by sunlight is responsible for most of marine
primary production. Absorption and scattering of phytoplankton control the light field (at
least for Case 1 waters which comprise a large majority of ocean waters), and thus, affect
the spectral reflectance of the ocean surface. Such spectral changes provide information
about ocean color (Gordon and Morel, 1983)

In all radiative transfer modeling of the marine environment the inherent optical
properties (IOPs) are either the necessary input parameters, or in the case of inverse
problems, the output of the calculations. This makes them important in marine optics
studies. The measurement of the attenuation coefficient is relatively easy; an attenuation
meter consists of a collimated light source and a collimated light detector at a known
distance. The only major problem in the attenuation measurement are photons from
outside sources or the ones reflected on the instrument scattered into the light beam,
causing an underestimation of the attenuation coefficient value. The detector has a finite
aperture so that some of the forward scattered photons will be counted as part of the
direct beam, also causing an underestimation. However, the measurement of the
components of the attenuation, the absorption and scattering coefficients, is inherently
much more difficult, and researchers have been striving to minimize the measurement
errors of these parameters for most of this century.

The central idea behind any absorption measurement is to project a beam of light through
an absorbing medium. If one could measure all of the unabsorbed light of the direct beam,
as well as all of the scattered light, the only light lost would be the absorbed light. Thus,
absorption meters tend to be arranged to collect as much of the scattered light as possible.
The absorption coefficient measured in any optical device has two major sources of error,
both of which are due to the fact that natural suspensions tend to scatter light as well as
absorb it. In practice, the scattered light traverses a longer path through the absorbing
medium and so is more likely to be absorbed (the path length amplification factor).
Secondly, not all of the scattered light is collected due to the geometry of the absorption
meter (the scattering error).

One of the choices to measure in situ absorption is the cylindrical reflection tube. Such a
tube needs to be be long enough to provide a sufficient optical path to measure the low
absorption values typical for Case1 waters below 580nm. Any photon scattered forward
off the instrument axis is, at least in theory, reflected by the tube walls until it reaches the
detector. An ideal instrument of this kind should also have a reflector inside of the source
end to collect the backscattered photons.

The first working prototypes of such a device were developed by
Zaneveld and co- workers (Zaneveld and Bartz, 1984; Zaneveld et
al., 1990). One of the greatest problems of making a practical
reflective tube is the reflectivity coefficient of the walls. A
Monte Carlo study of the performance of a reflective tube
absorption meter (Kirk, 1992) shows that the results quickly
deteriorate with reflectivity decreasing from 100\%. As it is
virtually impossible to produce perfect reflecting walls,
especially ones that would not deteriorate with prolonged use of
the instrument, the concept of a quartz glass tube surrounded by
air was proposed instead (Zaneveld et al., 1990).  Assuming smooth
surfaces of the tube, all photons encountering the wall at an
angle to the wall surface smaller than the critical angle,
$41^\circ$, must be internally reflected. Therefore, a clean
quartz reflection tube should collect all photons scattered in the
angular range between $0^\circ$ and $41^\circ$ (if multiple
scattering is neglected). However, the large losses of photons
scattered at angles above $41^\circ$ are the main theoretical
source of error for quartz tube absorption meters.

The Monte Carlo calculations by Kirk (1992) were conducted for a
prototype absorption tube with an almost parallel light beam
resembling the Wetlabs ac-9 absorption meter. The results showed
that the relative error of absorption is always positive and
increases linearly with the ratio of scattering to absorption. The
error increases with decreasing wall reflectance and acceptance
angle of the receiver. Another study by Hakvoort and Wouts (1994)
used a Lambertian light source. In this case the absorption error
decreases with the decreasing angle of photon acceptance. However,
early prototypes of HiStar, the new Wetlabs spectrophotometer, has
a diverging beam limited to $20^\circ$. The receiving end of the
tube is illustrated in Fig. 1. It consists of a Light Shaping
Diffuser (LSD) in front of a lens. The fiber transmits light to a
spectrometer. Such an arrangement results in a large receiving
area which through the use of the LSD and the lens translates into
the small acceptance angle needed by the spectrometer.

In this paper we take account of the reflections at both ends of the tube, that were
neglected by previous studies of this kind. We also extend previous results by considering
a more realistic arrangement, by introducing weighting functions that quantitatively show
the scattering error as a function of angle, and by providing calculations for some cases of
practical interest.

\section{CALCULATION SETUP}

The Monte-Carlo code used was adopted from the code written to determine the effects of
self-shading on an in-water upwelling irradiance meter (Piskozub 1994). This is a forward
Monte Carlo algorithm, meaning that the photons are traced in the forward direction
starting from the light source. An absorption event ends a photon's history. The low
values of the optical depth inside the absorption tube make this approach comparatively
efficient.

The absorption tube studied (called henceforth $\alpha$-TUBE) is a
cylindrical shell of inner radius r=0.006 m and length d=0.23 m
which corresponds to the dimensions of the HiStar quartz
reflective tube (see Fig. 1). Thickness of the quartz wall is
$\Delta$r=0.002 m. The indices of refraction used are 1.33 for
water and 1.41 for quartz.

The photons inside the $\alpha$-TUBE are traced along their three
dimensional trajectories. The scattering or absorption events
inside the volume of the water sample are defined by inherent
optical properties of the medium: the absorption coefficient of
the medium a, scattering coefficient b, and the scattering phase
function $\beta(\theta)$. Random numbers are used to choose
whether a given photon traveling from point (x,y,z) towards
direction $(\theta,\varphi)$ will reach the wall or the end of the
cylinder. Otherwise, it is assumed that the photon ends its
trajectory inside the liquid medium, and the type of the event
(scattering or absorption) is determined by comparing a random
number to the single scattering albedo $\omega_0$ value. All
random number used in the code are uniformly distributed in the
open interval (0, 1).

We define the Cartesian coordinate system such that the axis of
the tube is co-located with the z-axis, and the center of the
source end of the tube defines the origin of the coordinate
system. Therefore, $\theta$ is the angle between the photon
direction and the tube axis and $\varphi$ is the angle of
projection of the photon direction onto the x-y plane. In every
scattering event the new direction of the photon is chosen using
the relevant phase function.

\begin{figure}
\btdplot{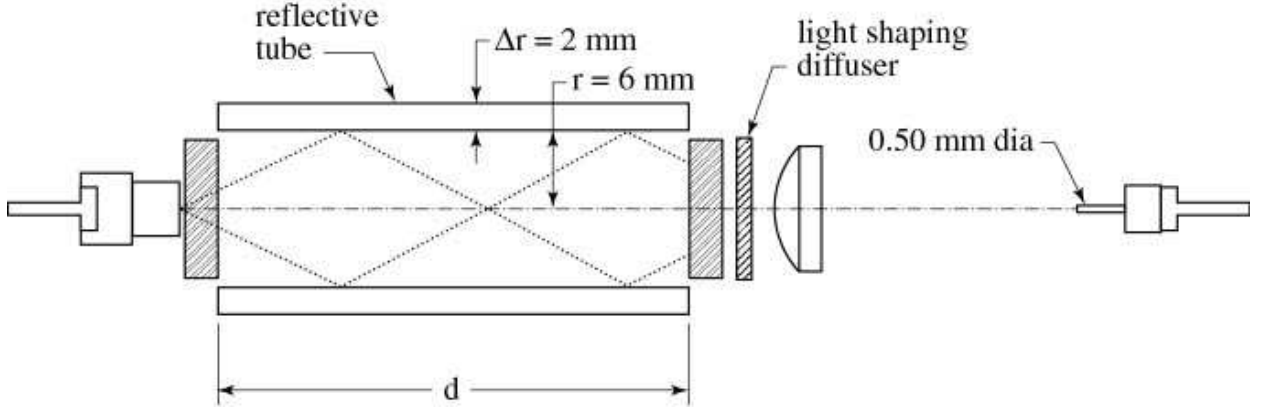}{1.} \caption{Schematic
representation of the reflective tube absorption meter setup
discussed in the paper}
\end{figure}

 The source of photons is assumed to be
within a circle of  the radius r = 1.5 mm emitting photons into a
cone of $25^\circ$ half-width. This represents the fiber head. The
angular distribution of photons is assumed to be Lambertian up to
the $25^\circ$. The position of the photon entering the tube
itself is calculated, taking into account the distance from the
fiber head to the tube entrance (0.019 m) and refraction from air
to water at the mouth of the reflective tube.

Such an $\alpha$-TUBE, defined above, is similar to the prototype
WET Labs HiStar, but it does not precisely model the detailed
radiance structure surrounding both the source and the receiver
ends of the reflective tube. We believe that those details are of
minor importance in comparison to neglecting the albedo of the
tube ends. Therefore we introduced the albedo of the source end
for photons returning to it from inside the tube, for which we
estimated the albedo as 0.3 (0.2 being the diffuse albedo and 0.1
the specular albedo). This is a rough assumption of the combined
effects of the input window and the silver colored metal surface
around the fiber head.

The quartz walls are treated as ideally smooth, reflecting and refracting photons according
to geometrical optics. The direction of a refracted photon is determined in the code by
Snell's law and probability of reflection and refraction by Fresnel's law. However, no
polarization effects are included. The photons are also followed inside the quartz, which
is assumed to be non-absorbing. All photons leaving the outer surface of the quartz tube
are treated as lost. The tube is assumed to be surrounded by an ideally black medium.

The receiver end of the tube is treated as a smooth surface
reflecting 8\% of the incident photons. This corresponds to the
angular average of reflection for the glass window and the Light
Shaping Diffuser (LSD) used. It is assumed that a photon hitting
the diffuser plate at an angle $\theta$ reaches the receiver with
a probability proportional to the angular diffusion distribution
(or scattering angle profile) of the LSD. The Light Shaping
Diffuser used in the instrument is a Physical Optics Corporation
$20^\circ$ LSD. Also, we use in the calculations two other
diffusers: $60^\circ$ LSD and $95^\circ$ LSDs. The angles are
half-widths of the angular diffusion function. This means that,
for example, a $60^\circ$ LSD has a 50\% transmission efficiency
at angle $30^\circ$. The angular diffusion distribution used was
determined by the least-square fits of the experimental curves
provided by the producer. The $20^\circ$ LSD was approximated by
two Gaussian functions (with all angles in degrees):
\begin{equation}
L_{20}(\theta)=A \exp\left[ -0.5 {\left( {\theta \over B }
\right)}^2 \right] + (1-A) \left[ -0.5 {\left( {\theta \over C}
\right)}^2 \right]
\end{equation}
where $A=0.85$, $B=8.0^\circ$, $C=15.3^\circ$. The $60^\circ$ LSD
was approximated by a single Gaussian function:
\begin{equation}
L_{60}(\theta)=A \exp \left[-0.5 {\left( {\theta \over B}
\right)}^2 \right]
\end{equation}
where $A=1.0$ and $B=24.5^\circ$. Finally, the  $95^\circ$ LSD was
approximated by a Gompertz function (Gompertz, 1825):
\begin{equation}
L_{95}(\theta) = A \exp\left[-\exp\left(-\left( {
{\theta-\theta_0} \over B} \right)\right)\right]
\end{equation}
where $A=1.0$, $B=7.33^\circ$ and $\theta_0=50.75$.

The shapes of the three LSD diffusion angle profiles are presented in Figure 2.
\begin{figure}
\btdplot{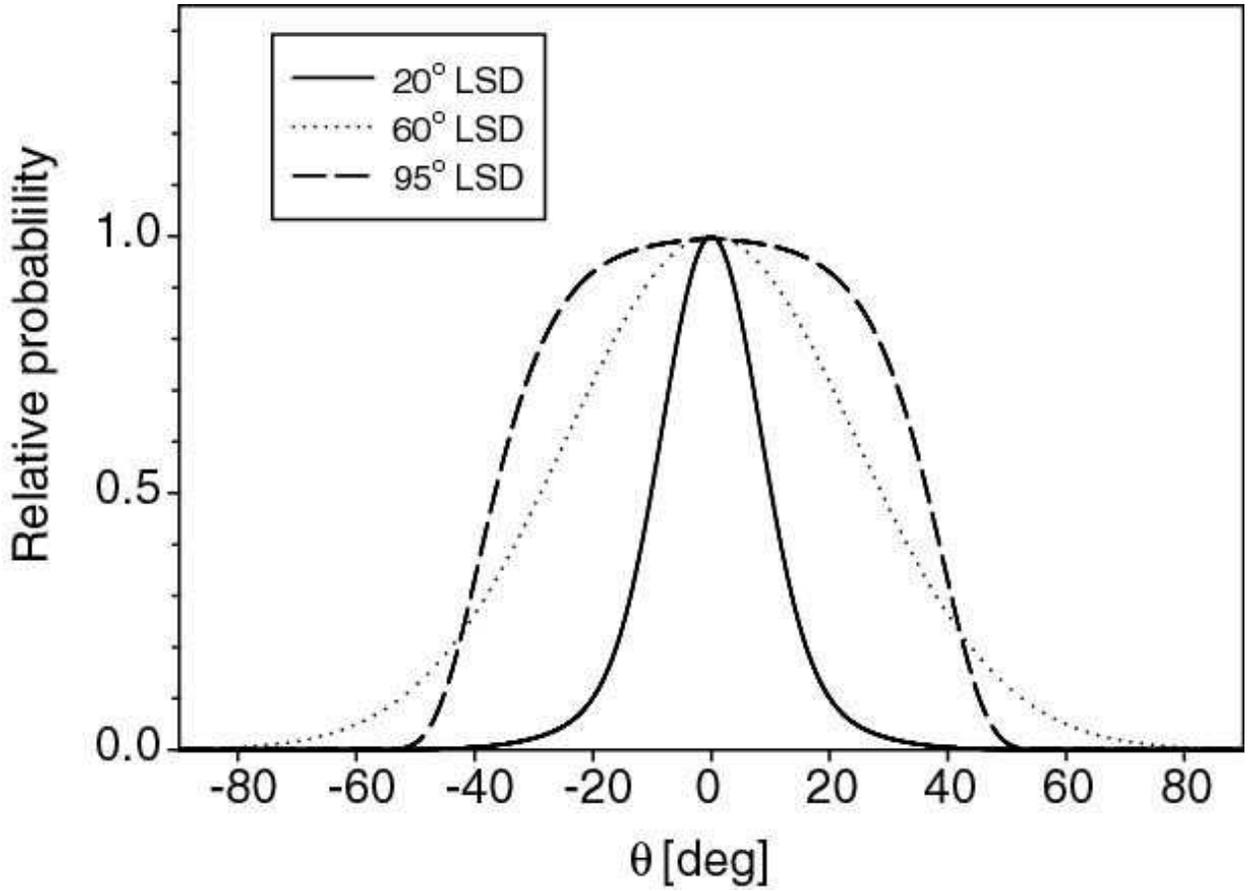}{1.} \caption{Light diffusing
characterisitics of the Light Shaping Diffusers for three
different types of diffusers.}
\end{figure}
The standard Petzold scattering phase function for open turbid
waters was used (Petzold 1972) except for the calculations of the
effect of the phase function shape on the measurement error where
Henyey-Greenstein phase functions (Henyey and Greenstein, 1941)
were used.

Each Monte Carlo calculation was performed twice: one to determine
how many photons are recorded if there is no absorption and
scattering ($P_0$) and the other for the IOPs (absorption,
scattering and phase function) of the aquatic medium being studied
(P). We define the measured absorption am as Kirk (1992):
\begin{equation}
a_m = (1/d) \ln (P_0/P)
\end{equation}
where d is the length of the cylinder.

It must be noted that the first program run corresponds to
calibration of the instrument in ``ideal water,'' not in air. We
use the water index of refraction for this medium so as not to
influence the new photon direction in refraction events.
Therefore, the results are different from what one would obtain if
the tube was filled with air. Such a non-scattering and
non-absorbing liquid medium does not exist. However, it is the
most obvious reference value to use in the calculations aimed at
studying the calibration of the instrument, because it does not
introduce any photon losses due to scattering (b=0) and does not
change the behavior of photons on the quartz-liquid border.

\section{RESULTS}

\subsection{SEMI-ANALYTICAL CONSIDERATIONS}
In this section we present qualitative arguments related to the
optical phenomena influencing the fate of photons in the
absorption tube. The purpose of this is to provide approximate
results that can be compared with exact Monte Carlo results later.
The semi- analytic results give a better understanding of the
underlying physics. If, for example, the IOPs of the studied water
sample are assumed to be: b=0.8 m$^{-1}$, a=0.2 m$^{-1}$, and the
Petzold San Diego Harbor (turbid water) phase function, one can
expect that 4.5\% ($\exp(- a d)$) of all photons will be absorbed
and 16.8\% ($\exp(-b d)$) will be scattered in the water volume
inside the quartz tube.

The phase function used, determines that 92.9\% of all scattering
events take place in the 0-41$^\circ$ range. Assuming for
simplicity, that before scattering, all photons traveled parallel
to the tube axis, all those scattered photons will be reflected
back into the tube on all encounters with the tube wall.
Similarly, all photons backscattered into the range of
139-180$^\circ$ stay in the tube on their way back to the source
end. These represent 0.3\% of all photons. The remaining 6.8\% of
all scattered photons are scattered into the range 41-139$^\circ$.
These photons have a large probability of leaving the tube on the
first encounter with the wall. The angles to the z-axis at which
those photons travel make it virtually certain that they will
leave the tube before reaching either end, unless they are
scattered close to $41^\circ$. For example at $\theta=45^\circ$, a
photon scattered in the very center of the tube will hit the wall
9 times before reaching the end of the tube making the probability
of not leaving it $<10^{-8}$. The 6.8\% of scattered photons that
are lost translates into 1.1\% of all traced photons being lost
through the walls.

After the absorption and scattering losses are taken into account,
94.3\% of all traced photons reach the receiver end of the tube.
Thus, the assumption of 8\% specular albedo of the receiver end of
the tube means that 7.5\% of all traced photons are reflected from
the receiver end of the tube. Absorption and scattering losses
result in only 7.1\% of all initial photons coming back to the
source end after being reflected. The backscattered photons, which
are attenuated in a pathlength approximately equal to the tube
length, increase this number to 7.2\%.

Another important aspect of a quartz absorption tube is the length of the path that the
photons travel in quartz instead of water. Any distance traveled in quartz decreases the
attenuation due to the negligible absorption coefficient in the visible range of the
spectrum (Kirk 1992). This decreases the measured absorption values because the
receiver is reached by more photons due to the shorter path through the absorbing
medium. In the studied tube geometry, the ratio of tube diameter to wall thickness is only
6:1, and considering that every time the photon goes across the water volume it needs to
cross the wall twice (out and back into the tube) it would seem that this effect must be
overwhelming. However, due to Snell's law and the narrow angles between the direction
of most photons and the wall, the average path of photons in quartz is much smaller.

\subsection{SANITY CHECKS OF THE MONTE CARLO RESULTS}

Even if rough approximations of the results are possible with
simple arithmetic calculations, Monte Carlo modeling includes all
optical phenomena taking place inside the tube. This can be
illustrated by comparing the approximate values from the previous
section to the results of a Monte Carlo run using 40 million
photons with the same IOPs. The number of photons lost through the
walls is 1.74\% (estimated above as 1.1\%). The discrepancy can be
explained by the divergence of the input beam: photons entering
the tube at an angle may leave it through the walls, even if
scattered from its original direction at less than the angle of
total internal reflection, resulting in higher scattering losses.

The Monte Carlo derived absorption losses percentage is 4.09\%
(semi-analytically estimated as 4.5\%). The reason for this is
that the path length of the photons is in quartz instead of in
water. The Monte Carlo code makes it possible, by tracing each
photon individually, to calculate the average path traveled by the
photons in water to be 21.15 cm and in quartz: 2.83cm. This means
that the path in water is 8\% shorter than the length of the
reflective tube, a value much higher than that calculated by Kirk
(1992) for a parallel beam in which only the scattered photons
encounter the walls. The losses by absorption on the front end of
the tube are 4.84\% of all photons (estimated as 5.0\%). The small
difference is the net result of smaller absorption losses and
greater scattering losses.

The effect of the shortened optical path, described above, is not very grave if the way the
instrument is calibrated is taken into account. All calibration techniques involve a
comparison of the received signal (that is the number of photons reaching the receiver)
for the measured sample with pure water or air values. Using pure water almost
completely removes the error due to non-scattered photons traveling partly inside the
wall, as the effect will be identical for water of any IOPs. Calibrating the instrument in air
leaves some error due to path length in quartz because air has a different index of
refraction making the path length in quartz shorter. There is, however, a much bigger
source of  error in calibrating the instrument without water, which is the removal of total
internal reflection on the quartz wall by surrounding both its sides by the same medium.

\subsection{VARIATION IN THE SCATTERING-ABSORPTION RATIO}

It was shown by Kirk (1992) that for a reflective tube absorption
meter propagating an almost parallel beam, that the ratio of the
measured to the true value of the absorption coefficient $a_m/a$
increases linearly with $b/a$ at a rate depending on the phase
function used. We decided to test whether such a relationship will
hold for the relatively divergent light beam and limited
acceptance angle of the $\alpha$-TUBE. Figure 3 shows that this is
indeed the case for all studied angles of acceptance (depending on
the LSD type used). The slope coefficients w of the linear fit
\begin{equation}
a_m = a + w b
\end{equation}
are 0.194 for $20^\circ$ LSD, 0.139 for $60^\circ$ LSD, and 0.0957
for $95^\circ$ LSD. These results suggest that the scattering
losses of the instrument increase with the decreasing angle of
acceptance of the receiver. It must be noted that the values of
the coefficient w depend on the scattering phase function used.

Figure 3 shows that there is a small offset in Equation 5.  Its
general version $a_m = a + w b +o$ (where o is the offset) can be
transformed to
\begin{equation}
a= { a_m - w c - o \over 1 - w}
\end{equation}
which, for a given phase function and absorbing tube geometry,
allows for determining the true a value if attenuation c is
measured independently. The parameters w and o may be calculated
for the given absorption meter geometry if the phase function is
known at least approximately. The values of the coefficient w, for
the physically important range of Henyey-Greenstein asymmetry
parameter g and the $20^\circ$ LSD receiver are shown in Figure 4.
\begin{figure} \btdplot{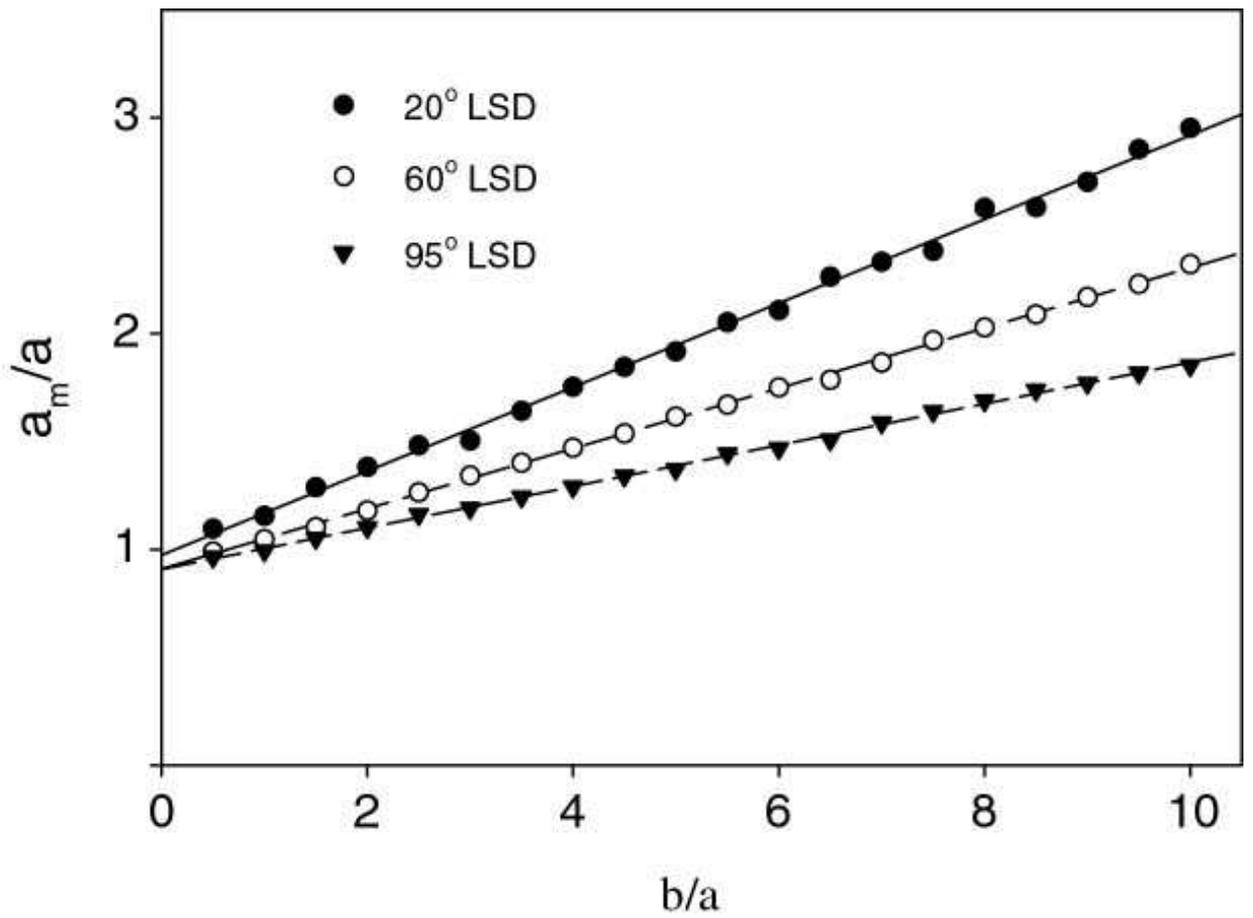}{1.} \caption{Ratio of the
measured to the true absorption coefficient $a_m/a$ as a function
of the scattering-absorption ratio $b/a$ and $a=0.2m^{-1}$,
Petzold "turbid" phase function for three different Light Shaping
Diffusers in front of the receiver lens.}
\end{figure}

\subsection{VARIATION IN THE SCATTERING PHASE FUNCTION}
The measurement error in absorption depends on the scattering
phase function of the aquatic medium (Kirk, 1992). To study the
effect of the scattering phase function on absorption measured by
the given instrument setup with $20^\circ$ LSD, we performed a
sequence of Monte Carlo calculations for the Henyey-Greenstein
phase function (Henyey and Greenstein 1941) by varying the
asymmetry parameter g
\begin{equation}
\beta(\theta)= {1-g^2 \over {(1+g^2 - 2g \cos\theta)}^{3/2} }
\end{equation}
where g is a parameter determining the shape of the phase
function. In this paper we assume range of g from $g=0$ (isotropic
scattering) to $g=1$ (forward scattering).
\begin{figure} \btdplot{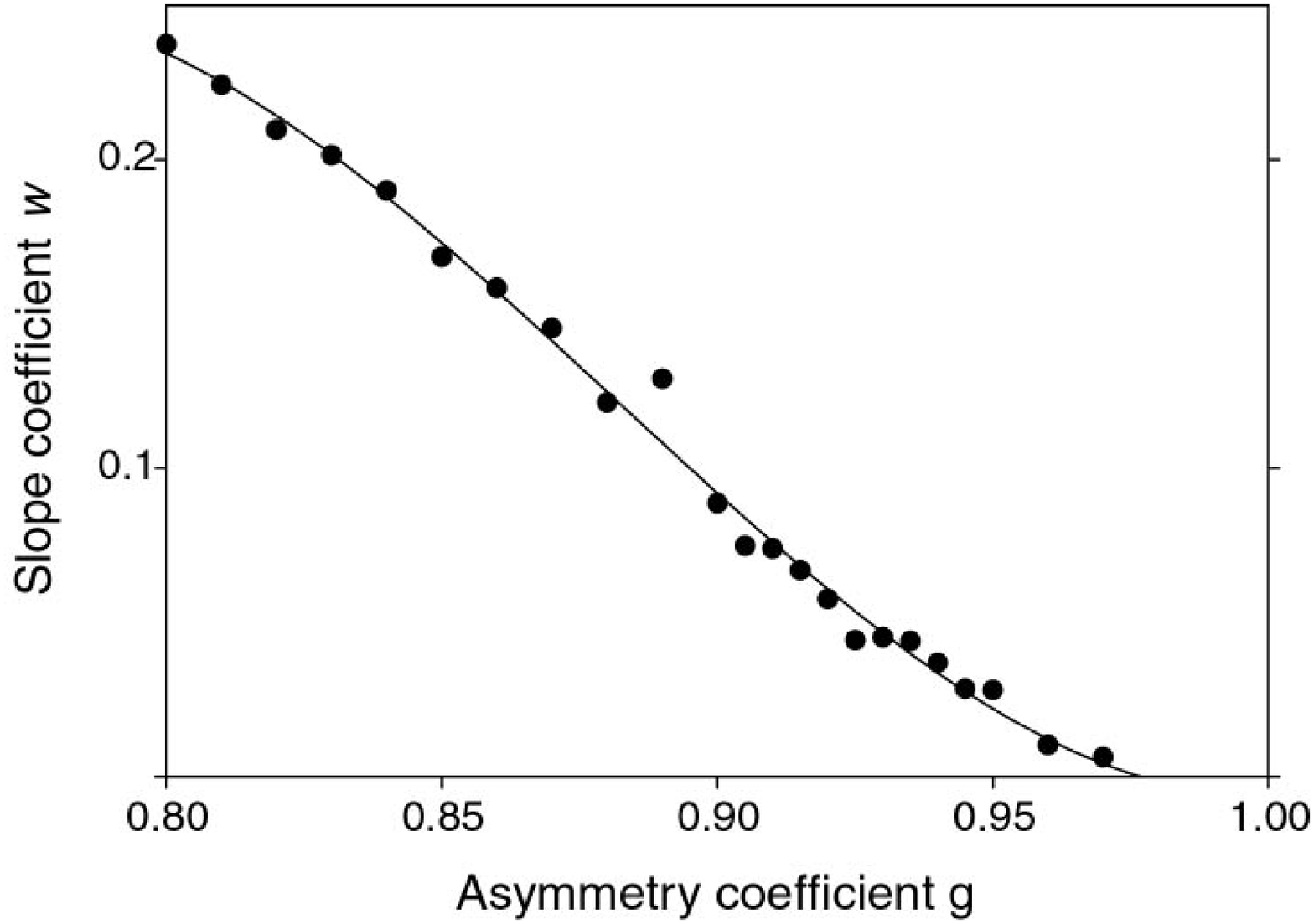}{1.} \caption{ The slope
coefficient w as a function of Henyey-Greenstein asymmetry factor
g for the $20^\circ$ LSD receiver geometry.}
\end{figure}
\begin{figure} \btdplot{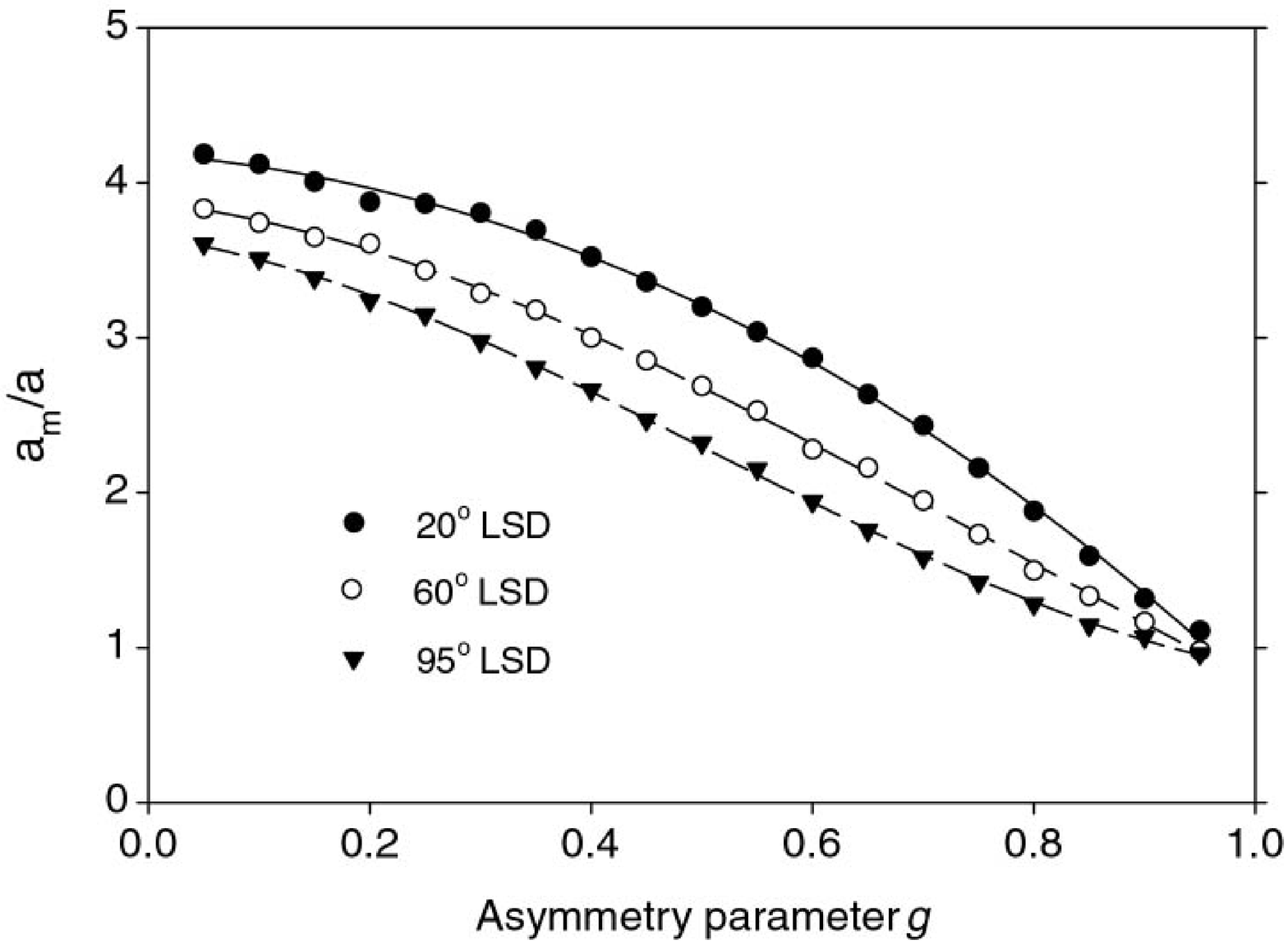}{1.} \caption{Ratio of the
measured to the true absorption coefficient $a_m/a$ as a function
of the Henyey-Greenstein scattering phase function asymmetry
parameter g; $a=0.2 m^{-1}$, $b=0.8 m^{-1}$, Petzold turbid phase
function, for three different Light Shaping Diffusers in front of
the receiver lens. Solid lines are third order polynomial fit.}
\end{figure}

The results in Fig. 5 show $a_m/a$ as a function of the asymmetry
parameter. They were obtained by running the code for the studied
tube equipped with three different LSD plates ($20^\circ$,
$60^\circ$, and $95^\circ$). As expected, for g approaching 1 the
$a_m/a$ ratio is close to 1 as all photons are scattered almost in
the direction that they were headed before the scattering. For a
more realistic range of g values between 0.8 and 0.9 (Mobley,
1994) the measured absorption increases when g decreases due to
the greater scattering losses of the photons. The value of $a_m/a
<1$ for $g=0.95$ and $95^\circ$ LSD is not a statistical error,
but the effect of the slightly longer path of scattered photons in
the quartz, in comparison to the non-scattering "ideal water"
which results in decrease of absorption.

For small g values the $a_m/a$ values (Fig. 5) approach the
maximum (a+b)/a at which all the scattered photons are lost and
the measured value of absorption equals the total attenuation. In
the case studied (a+b)/a = 5. However, even for an isotropic
scattering phase function (g=0) the value of the scattering error
does not reach this maximum value because some photons are
scattered at angles small enough to be recorded by the receiver.
Again, the figure shows that increasing the angle of acceptance of
the receiver decreases the absorption error for all realistic
phase functions.

\subsection{VARIATION IN THE ACCEPTANCE ANGLE}
Results presented by Kirk (1992) for a reflective tube absorption
meter with an almost parallel beam show a rapidly increasing value
of $a_m/a$ with the acceptance angle decreasing below $90^\circ$.
On the other hand, Hakvoort and Wouts (1994) suggest that for the
reflective tube with Lambertian light source and metallic
reflective walls, the measurement error increases with the
acceptance angle of the receiver. The Monte Carlo code was run for
several values of the receiver angle of acceptance to test if the
similiar situation arises for the  $\alpha$-TUBE. No Light Shaping
Diffuser was used in this case. The results are presented in
Figure 6.
\begin{figure} \btdplot{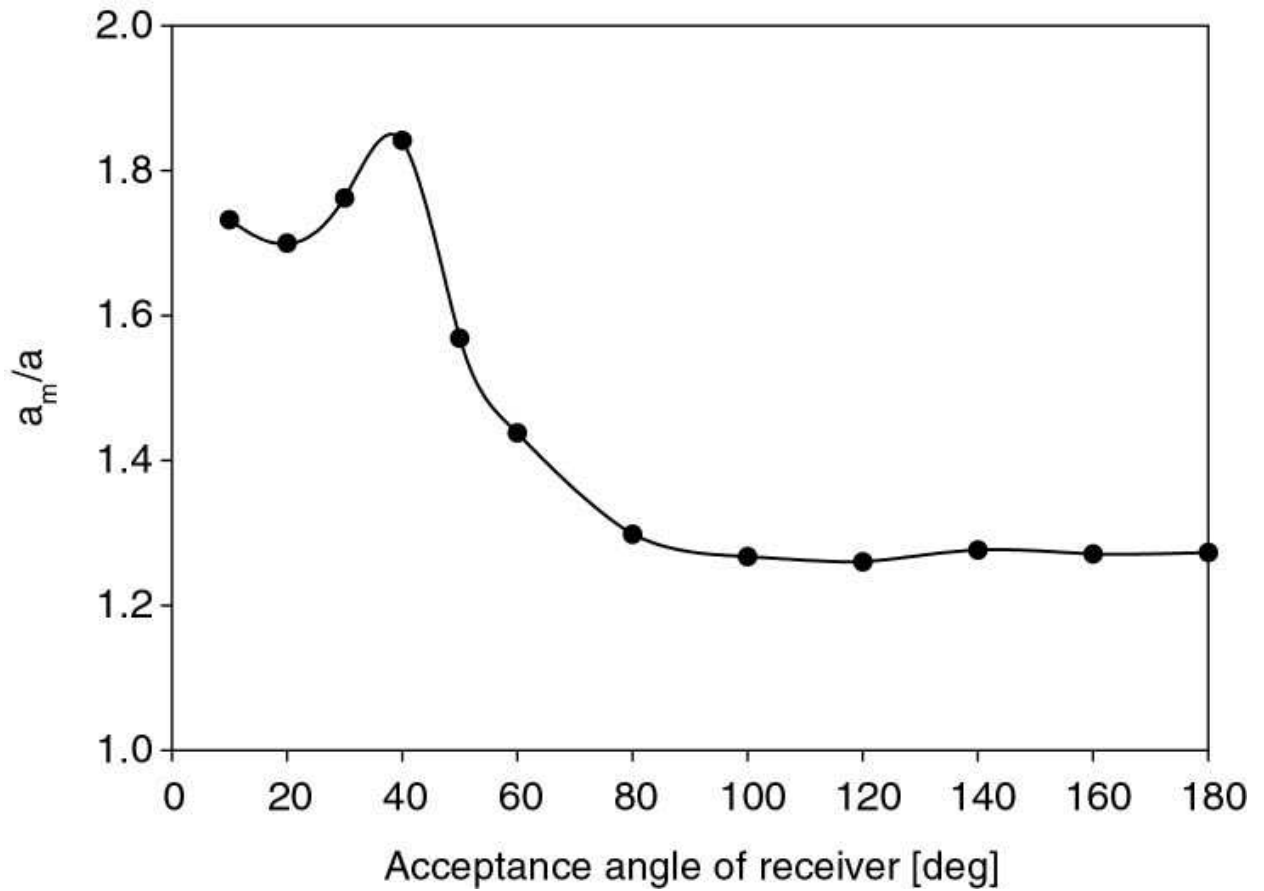}{1.} \caption{Ratio
of the measured to the true absorption coefficient $a_m/a$ as a
function of the acceptance angle of the receiver, $a=0.2 m^{-1}$,
$b=0.8 m^{-1}$, Petzold turbid phase function.}
\end{figure}

It can be seen that for large acceptance angles ($>40^\circ$)  the
$a_m/a$ values have a similar shape to that of a parallel beam
case. However, for acceptance angles in the $0-40^\circ$ range a
minimum can be observed, instead of a high peak as in the case of
a parallel beam. The reason for that is that the source beam is
more divergent than the angle of acceptance. For small acceptance
angles the receiver accepts some scattered photons that would not
be accepted if they reached the receiver without a scattering
event. This is because their angle of incidence is reduced by the
scattering event. This reduces the scattering error of the
absorption measurement. Unlike the Lambertian source case, there
is no major increase of $a_m/a$ values for large acceptance angles
(although a small increase above $120^\circ$ may be discerned). It
should be noted that the LSD plate used in the instrument
corresponds roughly to an acceptance angle of $20^\circ$. Although
this angle is close to a local minimum of $a_m/a$, it is still
within the range of the highest levels of absorption error.

\subsection{ANGULAR FUNCTION OF PHOTON LOSS PROBABILITY
$W(\theta)$}

Full understanding of the mechanism of scattering loss is not
possible without determining the angular relationship of the
scattering error of the reflective tube absorption meter. The
fraction of scattered light lost because of the absorption after
being scattered at a given angle can be defined as a loss function
$W(\theta)$. The angular integral of this function multiplied by
the phase function is the error of the measured absorption value
$\varepsilon = a_m-a$
\begin{equation}
\varepsilon=\int_0^\pi W(\theta) \beta(\theta) \sin(\theta)
d\theta
\end{equation}

The advantage of the W function is that it is not dependent on the
shape of the phase function but only on the tube geometry. This is
true only if multiple scattering is neglected because a second
scattering event influences the fate of a photon already
scattered. In order to differentiate among the possible sources of
the scattering loss error of the absorption value we decided to
calculate three variants of the W function (a) $W_0(\theta)$ for
scattering losses due to photons lost through the tube walls, (b)
$W_1(\theta)$ same as $W_0(\theta)$ and for scattering losses due
to additional photons absorbed by the source ("front") end of the
tube, (c) $W_2(\theta)$  same as $W_1(\theta)$  and for scattering
losses due to additional photons lost on the LSD.

By "additional photons" we mean that only the difference between
the number of photons lost by the absorption and scattering and
the "ideal" water is taken into account. In each of the three
cases the actual value of W for angle range is derived by dividing
the number of photons scattered into the angular sector and
subsequently lost by the sum of these photons plus the number of
photons scattered into the same sector which are subsequently
recorded by the receiver. Fig. 7 presents the W functions
calculated by running the code for 40 million photons with $a=0.2
m^{-1}$, $b=0.8 m^{-1}$, Petzold turbid phase function, and
$20^\circ$ LSD. In this case the number of multiple scattered
photons is 7\% of all which are scattered. That means that the
influence of the phase function shape on the W function is of
secondary importance. All three functions have the sigmoid shape
for the forward scattered photons between $0-90^\circ$. There is
no sharp step at $41^\circ$ and $139^\circ$ because the divergence
of the beam causes some photons to travel at an angle to the tube
axis before scattering which allows photons to escape through the
wall even if scattered at an angle smaller than the total internal
reflection angle. On the other hand, a photon scattered at more
than $41^\circ$ back towards the axis may survive the encounter
with the wall without being refracted out of the tube. The
symmetrical shape of $W_0$ is caused by the fact that losses on
the "front" end are not taken into account.  A photon
backscattered at angle $180-\phi$ has the same chance of being
lost through the walls as the one scattered at $\phi$. The
symmetry is broken only by the assumed diffusive albedo of the
light source end of the tube because some photons leave the tube
by the wall after being diffused on its source end.

Adding the photon losses on the light source end of the tube
(function $W_1$) changes the symmetry. Most of the backscattered
photons are lost in this way as well as some of the forward
scattered. The latter is caused by two reasons. Some
forward-scattered photons are reflected by the receiver end of the
tube back towards the source end.  However, it is even more
important that some of the forward scattered photons are absorbed
on the "front" end because they were scattered after they were
reflected on the receiver end. The reverse direction of those
photons before scattering complicates the $W(\theta)$ function,
but we decided to include these scattering events in the
statistics used to calculate the functions because they contribute
to the total scattering losses.

Receiver end losses are caused by more photons missing the
receiver due to reaching it at a wider angle after a scattering
event. This changes the picture by, paradoxically, decreasing the
photon losses for photons that were scattered in the forward
direction (function $W_2$). This means that scattering a photon by
a small angle increases its probability of being accepted by the
receiver compared to the average of non-scattered photons. This is
caused again by the source beam being wider than the acceptance
angle of the receiver. The discrepancy between $W_1$ and $W_2$,
especially at scattering angles close to 0, is also partly caused
by the inclusion of photons scattered on their way back to the
source end of the tube. One of its consequences is that the
population of photons scattered at close to 0 degree is different
from the population of non-scattered photons.
\begin{figure} \btdplot{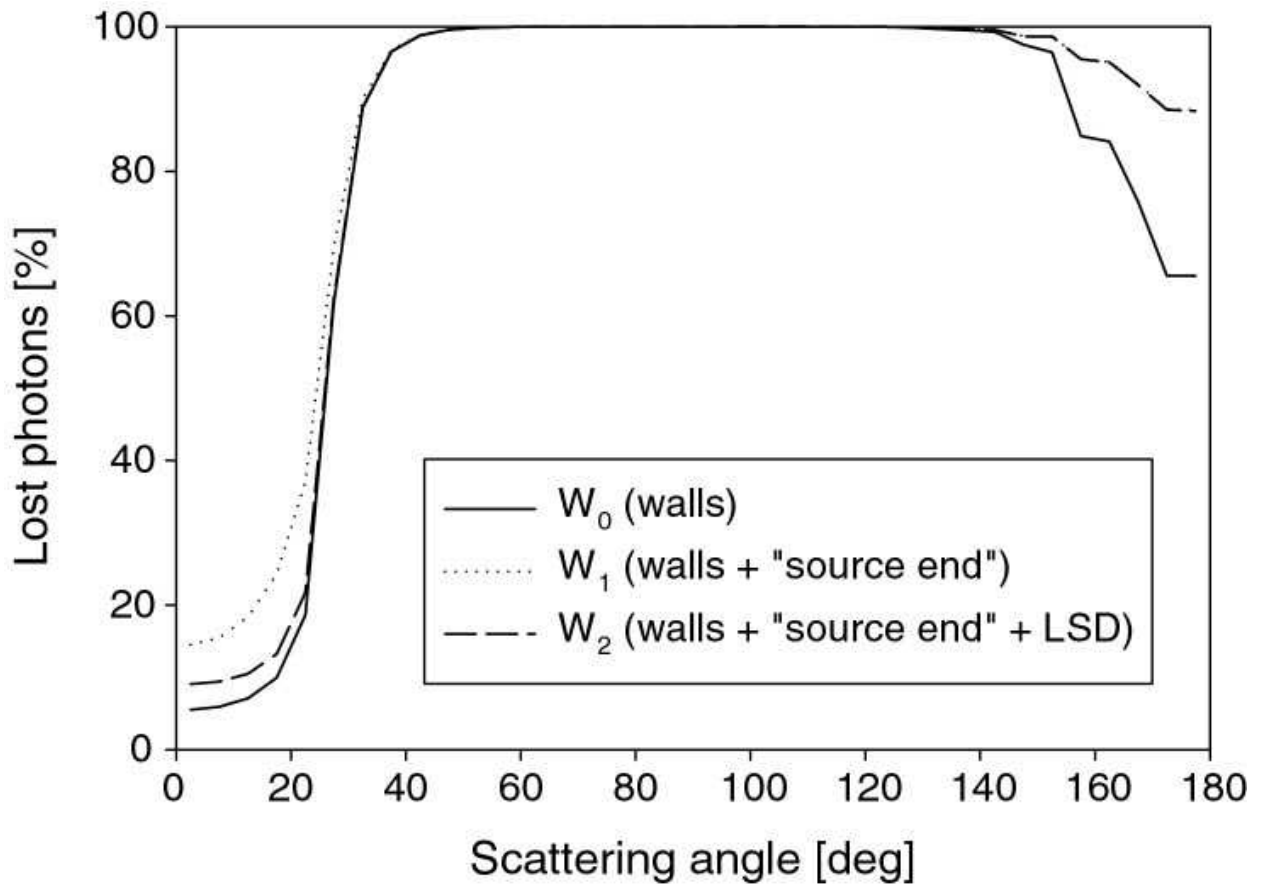}{1.} \caption{The photon
loss probability $W(\theta)$ as a function of the scattering
angle, $a=0.2 m^{-1}$, $b=0.8 m^{-1}$, Petzold turbid phase
function. These three functions represent (a) $W_0$ which defines
losses of photons leaving the tube through the quartz walls, (b)
$W_1$  same as $W_0$ and  losses on the light source end of the
cylinder, (c) $W_2$ which is the same as $W_1$ but with losses on
the receiver-end.}
\end{figure}

The shape of the W functions for angles greater than $90^\circ$ is
not important for the scattering error estimates because only a
fraction of all the photons is backscattered. For the $W_2$
function which  includes all the photon losses due to scattering,
the loss of backscattered  photons is close to 100\%. Therefore,
the best fit approximations to $W_0$, $W_1$ and $W_2$ were
calculated for the range $0-90^\circ$ only. The three functions
were fitted with a 4-parameter sigmoid

\begin{equation}
W(x)= y_0 + {A \over ( 1+ \exp(-(x-x_0)/B))}
\end{equation}
where for $W_0$ we have A = 0.935, $B = 2.53^\circ$, $x_0 =
26.72^\circ$, $y_0 = 0.061$, for $W_1$ we have  $A = 0.841$, $B =
3.35^\circ$, $x_0 = 25.75^\circ$, $y_0 = 0.157$, and for $W_2$ we
have $A = 0.901$, $B = 2.58^\circ$, $x_0 = 26.79^\circ$, $y_0 =
0.095$.

\section{CONCLUSIONS}

Monte Carlo calculations of the scattering error for the HiStar prototype absorption meter
show that the different blueprint of the instrument (more divergent light beam and a
limited angular view of the receiver) in comparison to other reflective tube absorption
meters does influence the error. Due to additional scattering losses by the view limited
receiver the absorption error is greater for all combinations of moderate optical
parameters. It is possible to correct the error by careful calibration, and the results
presented in this paper for the turbid water Petzold phase function are a step in this
direction. Independent measurements of attenuation may improve the error correction for
water samples of known (at least approximately) phase functions as shown in the paper.
However, the source of inter-instrumental discrepancy in absorption measurement that is
more difficult to correct is their different responses to phase function variability. One
possible solution is to use the photon loss function W. If the phase function of the studied
sea-water is known (or is possible to be estimated by comparing it to known phase
functions of similar water samples), the W function makes it possible to calculate directly
the value of the scattering error for the sample.

Our calculations show that the main source of the absorption error is the view limited
receiver. Therefore, we suggest that if technically possible (it would diminish the amount
of light collected by the receiver optical fiber) a diffuser of wider angular characteristics
should be used in the instrument receiver setup.

\section{REFERENCES}

\noindent Bricaud A.,  Babin M., Morel. A., and Claustre H., 1995:
"Variability in the chlorophyll- specific absorption coefficients
of natural phytoplankton: analysis and parameterization," J.
Geophys. Res. , 100,  13321--13332.
\newline

\noindent Gompertz B., 1925: "On the Nature of the Function
Expressive of the Law of Human Mortality," Phil. Trans. Roy. Soc.
London., 115, 513.
\newline

\noindent Gordon H. R. and Morel, A. 1983: "Remote assessment of
ocean color for interpretation of satellite visible imagery, a
review" in  Lecture notes on coastal and estuarine studies, vol.
4, Springler Verlag, New York, 114pp.
\newline

\noindent Hakvoort J.H.M, Wouts R. 1994: "Monte Carlo modelling of
the light field in reflective tube type absorption meter,"   Proc.
SPIE , 2258, Ocean Optics XII, 529--538.
\newline

\noindent Henyey, L. G., and J. L. Greenstein, 1941: "Diffuse
radiation in the galaxy,"  Astrophys. J., 93, 70--83.
\newline

\noindent Kirk, J.T.O, 1992: "Monte Carlo modeling of the
performance of a reflective tube absorption meter," Appl. Opt.,
31, 6463--6468
\newline

\noindent Mobley, C. D., 1994, Light and water: radiative transfer
in natural waters, Academic Press, San Diego.
\newline

\noindent Petzold, T. J., 1972: "Volume scattering functions for
selected ocean waters," SIO Ref. 72-78, Scripps Institution of
Oceanography, Univ. of California, San Diego.
\newline

\noindent Piskozub J., 1994: "Effects of surface waves and
sea-bottom on self-shading of in-water optical instruments" in
Ocean Optics XII, Proc. SPIE, 2258, 300--308.
\newline

\noindent Pope R.M.,  Fry E.S., 1997: "Absorption spectrum
(380-700 nm) of pure water. II. Integrating cavity measurements,"
Appl. Opt., 36, 8710--8723.
\newline

\noindent Zaneveld J.R.V, Bartz R, 1984: "Beam attenuation and
absorption meters" in Ocean Optics VII, M.A. Blizard ed., Proc.
SPIE, 489,  318--324.
\newline

\noindent Zaneveld J.R.V, Bartz R., Kitchen J.C, 1990: "A
reflective tube absorption meter," in Ocean Optics X, R.W. Spinard
ed., Proc. SPIE, 1302, 124-136.

\end{document}